\documentclass[twocolumn,pra]{revtex4}

\usepackage{graphicx}
\usepackage{color}
\usepackage{ulem}

\begin{document}
\title{Temporal profile of biphotons generated from a hot atomic vapor and spectrum of electromagnetically induced transparency}

\author{
Shih-Si Hsiao,$^{1,}$\footnote{Electronic address: {\tt shihsi.hsiao@gmail.com}}
Wei-Kai Huang,$^1$
Yi-Min Lin,$^1$
Jia-Mou Chen,$^1$ 
Chia-Yu Hsu,$^1$ and
Ite A. Yu$^{1,2,}$}\email{yu@phys.nthu.edu.tw}

\address{$^1$Department of Physics, National Tsing Hua University, Hsinchu 30013, Taiwan \\
$^2$Center for Quantum Technology, Hsinchu 30013, Taiwan
}

\begin{abstract}
We systematically studied the temporal profile of biphotons, i.e., pairs of time-correlated single photons, generated from a hot atomic vapor via the spontaneous four-wave mixing process. The measured temporal width of biphoton wave packet or two-photon correlation function against the coupling power was varied from about 70 to 580 ns. We derived an analytical expression of the biphoton's spectral profile in the Doppler-broadened medium. The analytical expression reveals that the spectral profile is mainly determined by the effect of electromagnetically induced transparency (EIT), and behaves like a Lorentzian function with a linewidth approximately equal to the EIT linewidth. Consequently, the biphoton's temporal profile influenced by the Doppler broadening is an exponential-decay function, which was consistent with the experimental data. Employing a weak input probe field of classical light, we further measured the EIT spectra under the same experimental conditions as those in the biphoton measurements. The theoretical predictions of the biphoton wave packets calculated with the parameters determined by the classical-light EIT spectra are consistent with the experimental data. The consistency demonstrates that in the Doppler-broadened medium, the classical-light EIT spectrum is a good indicator for the biphoton's temporal profile. Besides, the measured biphoton's temporal widths well approximated to the predictions of the analytical formula based on the biphoton's EIT effect. This study provides an analytical way to quantitatively understand the biphoton's spectral and temporal profiles in the Doppler-broadened medium.
\end{abstract}

\maketitle

\newcommand{\FigOne}{
	\begin{figure}[t]
	\center{\includegraphics[width=60mm]{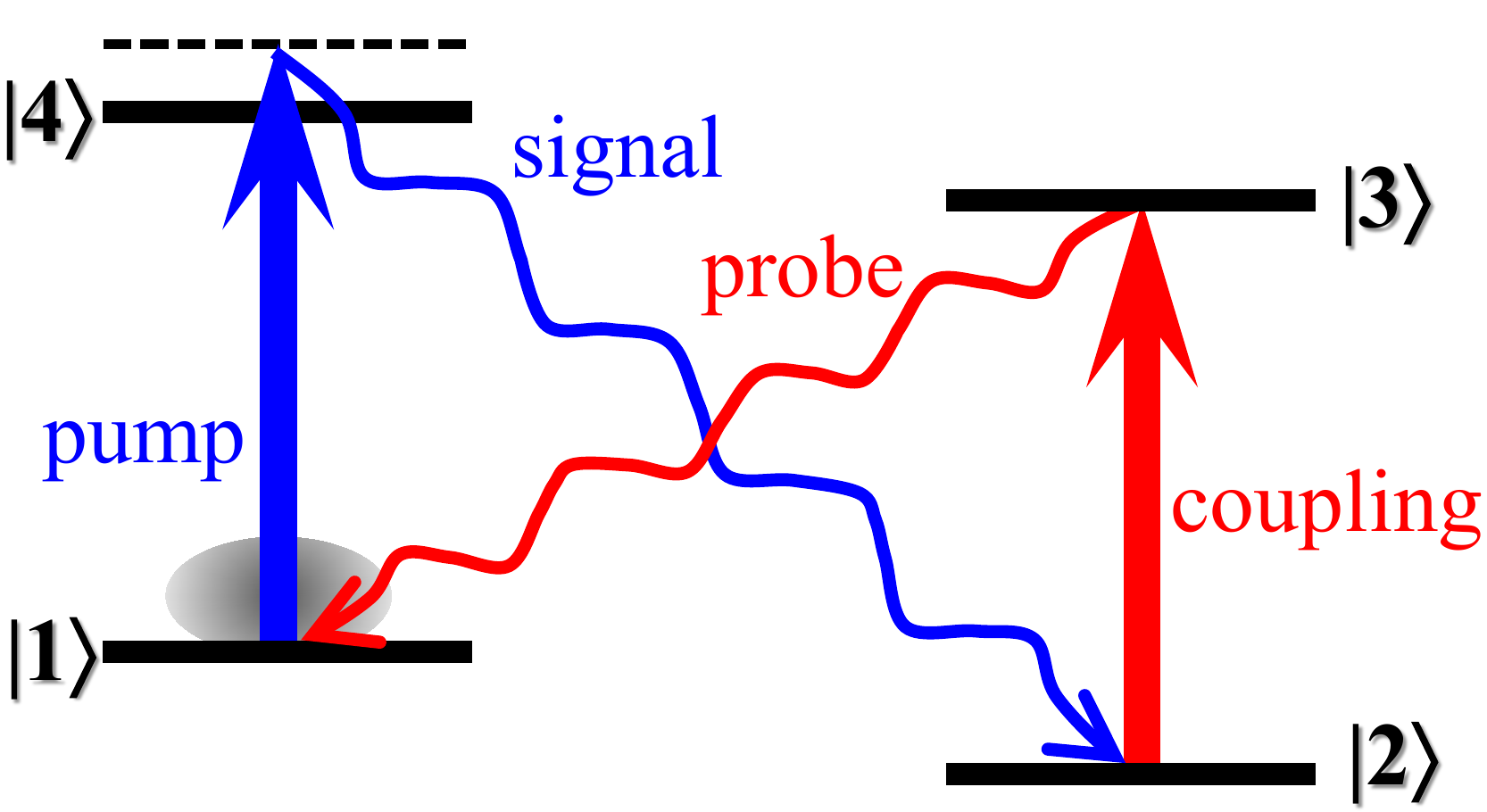}}
	\caption{
Transition diagram of the SFWM process. Under the presence of the pump and coupling fields, the vacuum fluctuation induces the generation of a pair of the signal and probe single photons. The pump transition has a large magnitude of the one-photon detuning, $\Delta_p$, such that the excitation to state $|4\rangle$ is negligible. In the experiment, a hot vapor of $^{87}$Rb atoms were utilized. States $|1\rangle$, $|2\rangle$, $|3\rangle$, and $|4\rangle$ are $|5S_{1/2},F=2\rangle$, $|5S_{1/2},F=1\rangle$, $|5P_{1/2},F=2\rangle$, and $|5P_{3/2},F=1,2\rangle$, respectively, and $\Delta_p$ = 2.0~GHz. Since the energy level of $|1\rangle$ is higher than that of $|2\rangle$, the signal (or probe) photon is also called the anti-Stokes (or Stokes) photon. 
}
	\label{fig:transition_diagram}
	\end{figure}
}
\newcommand{\FigTwo}{
	\begin{figure}[t]
	\center{\includegraphics[width=60mm]{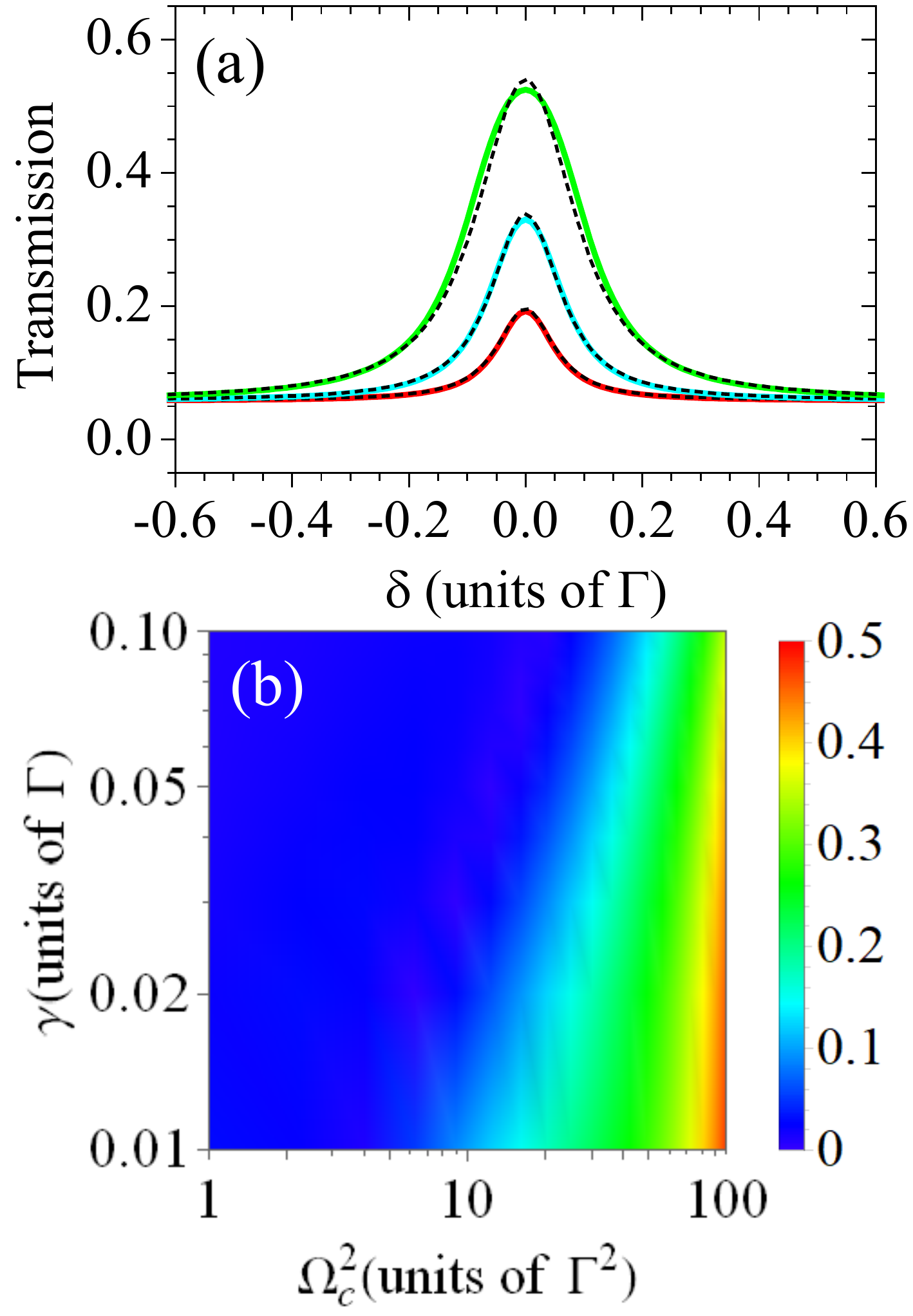}}
	\caption{
(a) Theoretical predictions of the biphoton's EIT spectrum of a Doppler-broadened medium. Solid lines are the numerical results calculated from Eq.~(\ref{eq:T_exact}), and dashed lines are the results of the analytical formula in Eq.~(\ref{eq:T_approximation}). Red, blue and green correspond to $\Omega_c =$ 2.5$\Gamma$, 3.5$\Gamma$, and 5.0$\Gamma$. The other relevant calculation parameters are $\gamma =$ 0.05$\Gamma$,  $\alpha_s=350$, and $\Gamma_D =$ 54$\Gamma$. (b) Percentage difference between the FWHM of the biphoton's EIT spectrum given by the numerical calculation and that given by the analytical formula as a function of $\Omega_c^2$ and $\gamma$.
}
	\label{fig:eit_simulation}
	\end{figure}
}
\newcommand{\FigThree}{
	\begin{figure}[t]
	\center{\includegraphics[width=60mm]{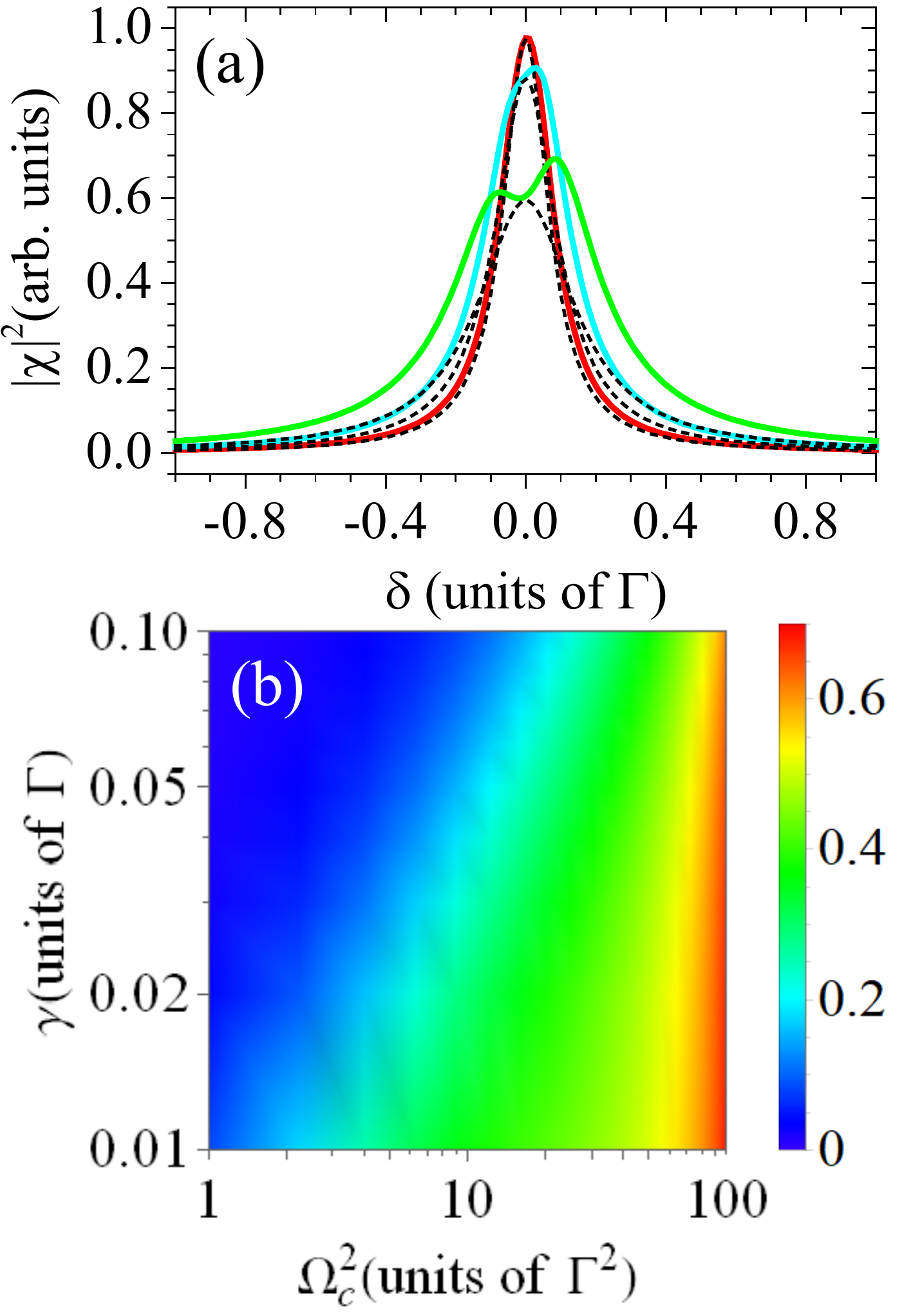}}
	\caption{
(a) Theoretical predictions of the biphoton's FWM spectrum of a Doppler-broadened medium. Solid lines are the numerical results calculated from Eq.~(\ref{eq:FWM_part}) and dashed lines are the results of the analytical formula in Eq.~(\ref{eq:FWM_part_th}). Red, blue and green correspond to $\Omega_c =$ 2.5$\Gamma$, 3.5$\Gamma$, and 5.0$\Gamma$. The other relevant calculation parameters are $\gamma =$ 0.05$\Gamma$ and $\Gamma_D =$ 54$\Gamma$. (b) Percentage difference between the FWHM of the biphoton's FWM spectrum given by the numerical result and that given by the analytical formula as a function of $\Omega_c^2$ and $\gamma$.
}
	\label{fig:fwm_simulation}
	\end{figure}
}
\newcommand{\FigFour}{
	\begin{figure}[t]
	\center{\includegraphics[width=60mm]{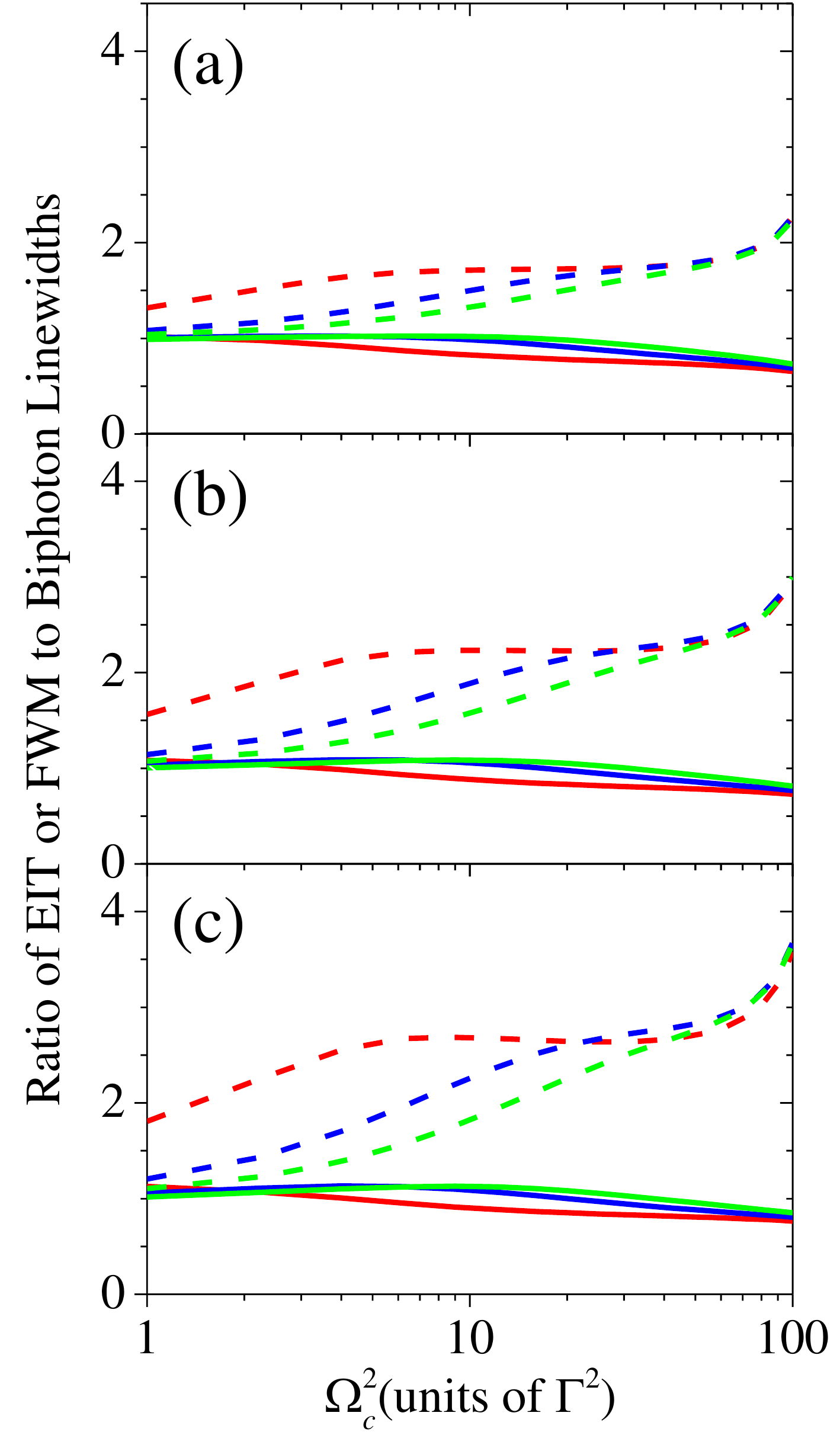}}
	\caption{
Comparisons between the linewidths of biphoton's EIT, FWM, and overall spectra, i.e., $\Gamma_{\rm EIT}$, $\Gamma_{\rm FWM}$, and $\Gamma_{\rm BI}$, respectively. The values of $\Gamma_{\rm BI}$, $\Gamma_{\rm EIT}$, and $\Gamma_{\rm FWM}$ are numerically calculated from Eqs.~(\ref{eq:biphoton_frequency}), (\ref{eq:T_exact}), and (\ref{eq:FWM_part}) with $\Gamma_D =$ 54$\Gamma$. Solid lines are the ratios of $\Gamma_{\rm EIT}$ to $\Gamma_{\rm BI}$, and dashed lines are those of $\Gamma_{\rm FWM}$ to $\Gamma_{\rm BI}$. Red, blue, and green colors represent $\gamma =$ 0.01$\Gamma$, 0.05$\Gamma$, and 0.1$\Gamma$. Subfigures (a), (b), and (c) correspond to $\alpha_s =$ 200, 350, and 500.
}
	\label{fig:three_linewidth_comparison}
	\end{figure}

}
\newcommand{\FigFive}{
	\begin{figure}[t]
	\center{\includegraphics[width=60mm]{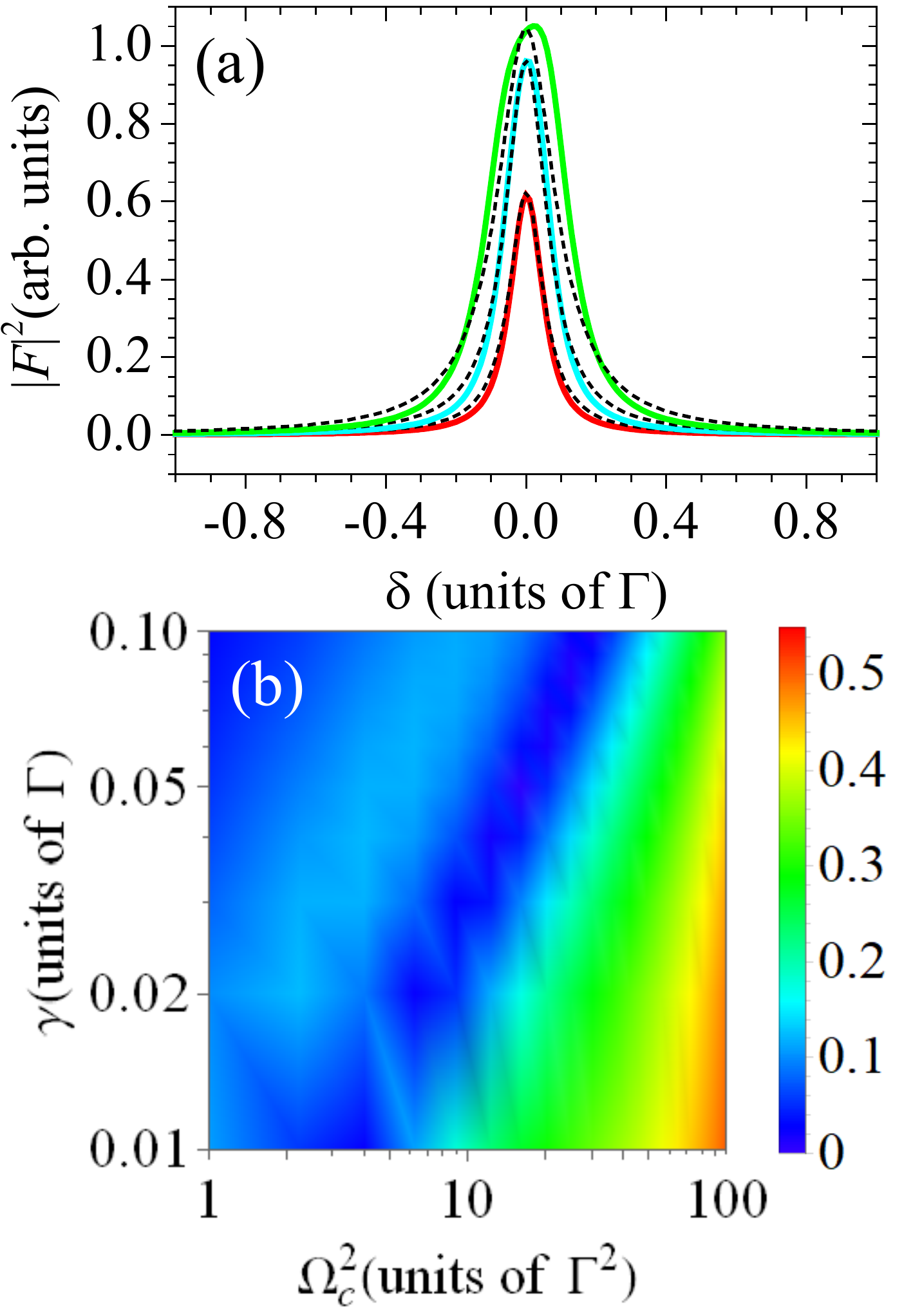}}
	\caption{
Theoretical predictions of the biphoton's overall spectrum of a Doppler-broadened medium. Solid lines are the numerical results given by Eq.~(\ref{eq:biphoton_frequency}), and dashed lines are the analytical results given by the product of Eq.~(\ref{eq:T_approximation2}). Red, blue and green correspond to $\Omega_c =$ 2.5$\Gamma$, 3.5$\Gamma$, and 5.0$\Gamma$. The other relevant calculation parameters are $\gamma =$ 0.05$\Gamma$, $\alpha_s=350$, and $\Gamma_D =$ 54$\Gamma$. (b) Percentage difference between the FWHM of the biphoton's overall spectrum given by the numerical result and that given by the analytical formula, i.e., $\Gamma_{\rm BI}$, as a function of $\Omega_c^2$ and $\gamma$.
}
	\label{fig:biphoton_simulation}
	\end{figure}
}
\newcommand{\FigSix}{
	\begin{figure}[!t]
	\center{\includegraphics[width=85mm]{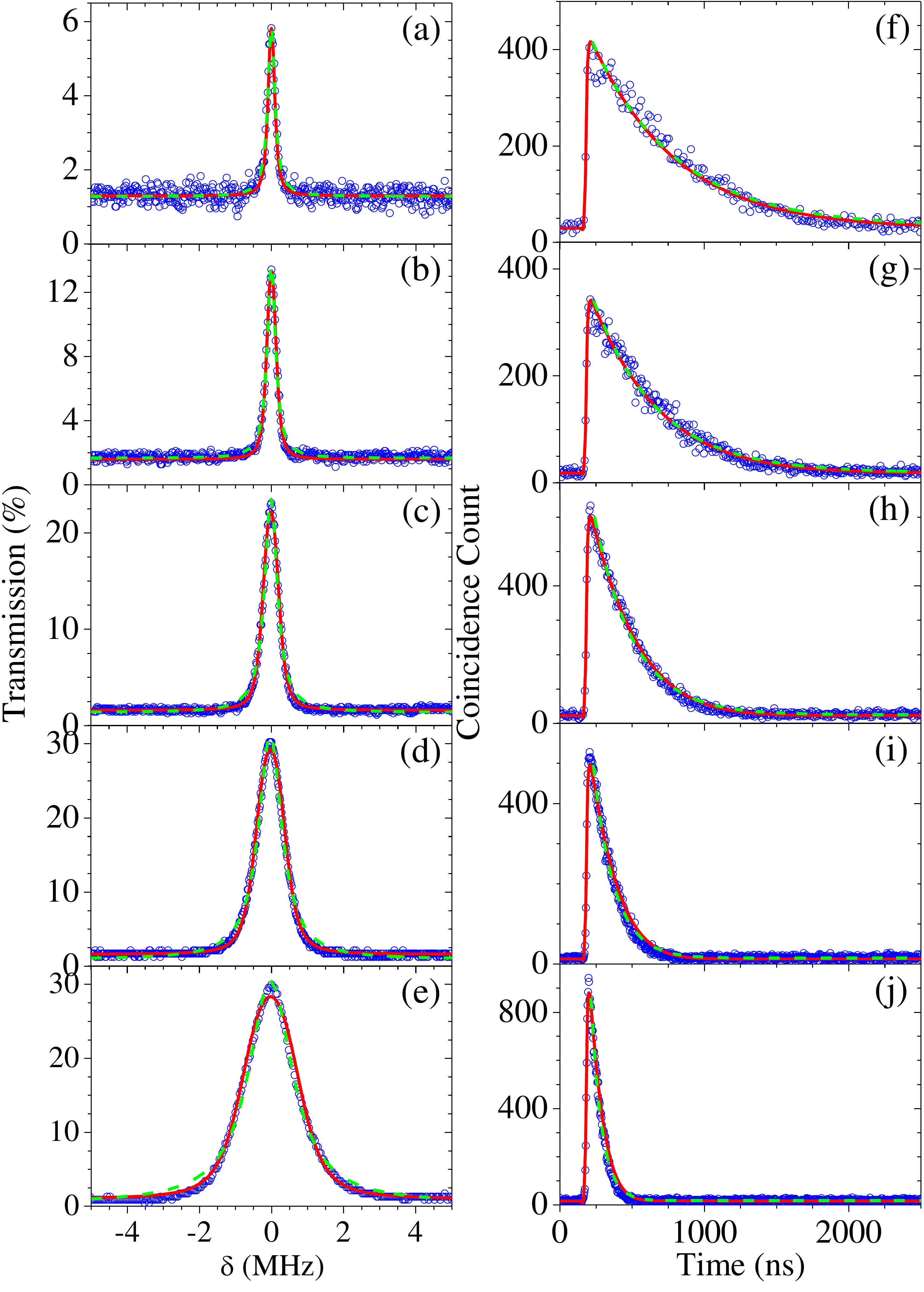}}
	\caption{
(a)-(e) From top to bottom, blue lines are the measured EIT spectra at the coupling powers of 0.5, 1, 2, 4, and 8 mW, and red lines are the theoretical predictions calculated from Eq.~(\ref{eq:T_exact}) with $\alpha_s  =$ 360 and $(\Omega_c, \gamma) =$ (1.75$\Gamma$, 0.020$\Gamma$), ($\sqrt{2}$$\times$1.75$\Gamma$, 0.022$\Gamma$), (2$\times$1.75$\Gamma$, 0.029$\Gamma$), (2$\sqrt{2}$$\times$1.75$\Gamma$, 0.044$\Gamma$), and (4$\times$1.75$\Gamma$, 0.086$\Gamma$). Green lines are the best fits of Lorentzian functions. (f)-(j) Blue lines are the experimental data of biphoton wave packets or two-photon correlation function at the same coupling powers as those in (a)-(e), respectively. 
The accumulation time of data in each figure was 240~s. The time-bin widths of the data from top to bottom were 12.8, 6.4, 6.4, 3.2, and 3.2~ns. Red lines represent the theoretical predictions calculated from Eq.~(\ref{eq:biphoton}). The calculation parameters of $\Omega_c$, $\alpha_s$, and $\gamma$ are set to the values determined in (a)-(e). Green dashed lines are the best fits of exponential-decay functions, indicating that the $e^{-1}$ time constants of the biphoton data from top to bottom are 580, 445, 270, 150, and 74~ns.
}
    \label{fig:experiment_data}
	\end{figure}
}
\newcommand{\FigSeven}{
	\begin{figure}[t]
	\center{\includegraphics[width=60mm]{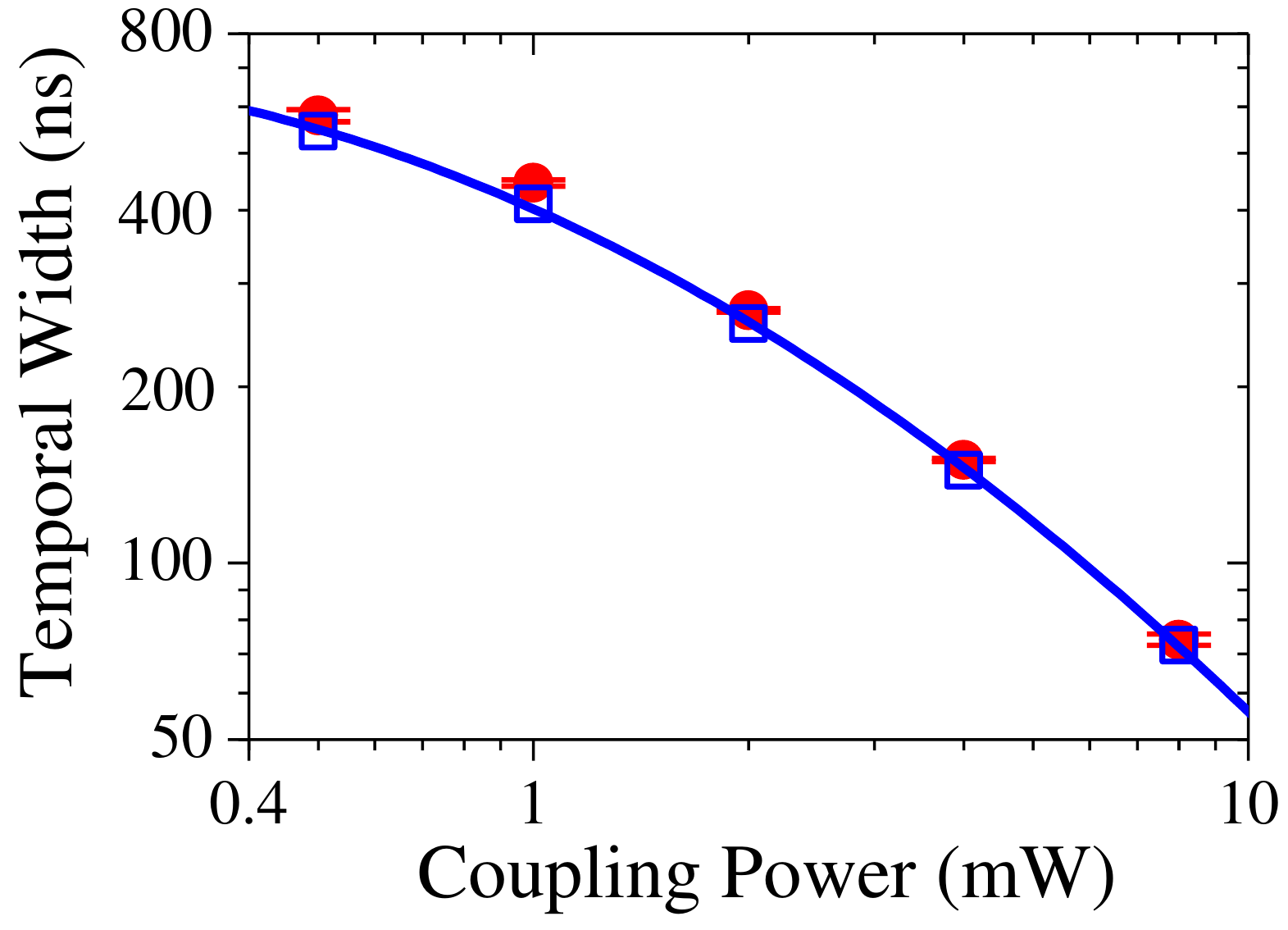}}
	\caption{
Temporal width of the biphoton wave packet as a function of the coupling power. Red circles are obtained from the best fits of the biphoton wave packets as the examples shown in Figs.~\ref{fig:experiment_data}(f)-\ref{fig:experiment_data}(j). Blue squares are the reciprocals of the analytical formula in Eq.~(\ref{eq:Gamma_EIT}) with the calculation parameters determined in Figs.~\ref{fig:experiment_data}(a)-\ref{fig:experiment_data}(e). Blue line is a smooth curve to guide eyes.
}
    \label{fig:eit_biphoton_linewidth}
	\end{figure}
}
\section{Introduction} \label{sec:introduction}
The biphoton is a pair of time-correlated single photons. When the first photon is detected to trigger a quantum operation, the second photon in the same pair will be employed in the quantum operation. The first and second photons can be regarded as the heralding and heralded single photons. Biphotons can be used in quantum information processing~\cite{QI1,QI2,QI3,QI4,QI5,QI6}, such as quantum communication~\cite{Q_communication_1,Q_communication_2,Q_communication_3}, quantum memory~\cite{Q_memory_1,Q_memory_2,Q_memory_3,Q_memory_4}, and quantum interference~\cite{Q_interference_1,Q_interference_2,Q_interference_3}.

One of the mechanisms to produce biphotons is the spontaneous four-wave mixing (SFWM) process~\cite{SWFM1,SWFM2,SWFM3,SWFM4,SWFM5,SWFM6,SWFM7,SWFM8,SWFM9,SWFM10,SWFM11,SWFM12,SWFM13,SWFM14,SWFM15,SWFM16,SWFM17,SWFM18,SWFM19, SWFM20,SWFM21,SWFM22,SWFM23,SWFM24}, which has been commonly employed with media of cold or hot atomic vapors. There are several theoretical models to study the temporal profile of the biphoton. In Ref.~\cite{SWFM14}, S.~Du {\it et al}. developed the theory of the time-correlation function of biphotons in a cold atomic vapor. The authors systematically studied the two-photon correlation function, i.e., the biphoton wave packet, discussed the properties of each term in wave packet, and predicted the effects of the dephasing rate and propagation delay time on the temporal profile of the biphoton wave packet. Based on the theoretical model in Ref.~\cite{SWFM14}, several groups conducted experiments to study the temporal profiles of biphotons with cold atoms~\cite{SWFM7,SWFM8, SWFM9,SWFM10,ColdAtoms1,ColdAtoms2}. Moreover, in Ref.~\cite{SWFM8} and \cite{SWFM14} the authors derive the analytical expressions of time correlated function, providing the convenient method to expected the temporal profile of biphoton in cold atom system.

The SFWM-generated biphoton source of a room-temperature or hot atomic vapor has the merits of adjustable linewidth, tunable frequency, and high generation rate. In Ref.~\cite{SWFM18}, C.~Shu {\it et al}. reported the generation of biphotons from a hot atomic vapor, and provided the numerical simulation for the time-correlation function of biphotons. After Ref.~\cite{SWFM18}, several groups improved the efficiency of biphoton generation with hot-atom sources, and compared data with results of numerical simulations~\cite{SWFM23, SWFM24, HotAtom1}.
However, there is no analytical formula available for the temporal profile of biphoton's time-correlation function or wave packet in the Doppler-broadened medium so far, and the systematic study on the temporal profile has not been reported before.

In this article, we report the systematic study on the temporal profile of the SFWM biphotons generated from a Doppler-broadened medium. Although the spectral profile of the biphoton is influenced by the electromagnetically induced (EIT) effect, the four-wave mixing (FWM) process, the phase matching condition, and the Doppler effect as well as the Boltzmann distribution of the velocity, we derived an analytical expression to show that the biphoton's overall spectrum is approximately Lorentzian. Thus, the temporal profile of the biphoton wave packet behaves like an exponential-decay function with the time constant, which is mainly determined by the EIT effect. Furthermore, we experimentally demonstrated that the measured biphoton's temporal profile is predominately characterized by the measured EIT spectrum. The predictions of the biphoton's temporal width calculated from the analytical formula are in good agreement with the experimental data.

We organize the article as follows. In Sec.~II, we start with the time-correlation function or the wave packet of the SFWM-generated biphoton in a Doppler-broadened medium, and derive the analytical expressions or formulas of the biphoton's spectral and temporal profiles. In Sec.~III, we describe our experimental setup for the measurement of the EIT spectrum with classical light and that of the biphoton wave packet. In Sec.~IV, we presented the data of the EIT spectra and the biphoton wave packets, and compare them with the predictions from the theoretical model. Finally, we give a conclusion in Sec.~V.

\section{Theoretical Predictions}
Biphotons or a pair of single photons are produced from a hot atomic vapor with the spontaneous four-wave mixing process (SFWM). Figure~\ref{fig:transition_diagram} shows the relevant energy levels and transitions of SFWM process for the generation of biphotons. All population is placed in the ground state of $|1\rangle$. The pump field is far detuned from the transition of $|1\rangle$ $\rightarrow$ $|4\rangle$, and the coupling field drive the transition of $|2\rangle$ $\rightarrow$ $|3\rangle$ resonantly. Due to the vacuum fluctuation, a pair of anti-Stokes and Stokes photons can be spontaneously emitted. The frequencies of the pump field, anti-Stokes photon, coupling field, and Stokes photon form the resonant four-photon transition. Note that, we employ the hyperfine optical pumping (HOP) field to empty the population in $|2\rangle$, which may contribute some residual light of the HOP field in its hollow region. The anti-Stokes photon propagates with the speed of light in vacuum since the pump field and anti-Stokes form the Raman process with the large one-photon detuning. The Stokes photon is slow light due to the effect of electromagnetically induced transparency (EIT) in two-photon on resonance of the coupling field and Stokes photon.

\FigOne

The temporal profile of biphoton wave packet is determined by the theory of the time correlation function between the anti-Stokes and Stokes photons. Considering phase-match condition, the time correlation function is shown below \cite{SWFM14}.
\begin{eqnarray}
	G^{(2)}(\tau) =  
		\left| \int_{-\infty}^{\infty} d\delta \frac{1}{2\pi} e^{-i\delta\tau} F(\delta) \right|^2,
\label{eq:biphoton}
\end{eqnarray} 
\begin{eqnarray}
	F(\delta) =  
		\frac{\sqrt{k_{as} k_s}L}{2} \chi(\delta)  
		{\rm sinc}  \left[ \frac{k_s L}{4} \xi(\delta) \right] 
		e^{i(k_s L/4) \xi(\delta)} ,
\label{eq:biphoton_frequency}
\end{eqnarray} 
where $\delta$ is the two-photon detuning between the anti-Stokes photon and pump field (or $-\delta$ is that between the Stokes photon and coupling field), $\tau$ is the delay time of the Stokes photon, $k_{as}$ and $k_s$ are the wave vectors of the two photons, $L$ is the medium length, $\chi(\delta)$ is the cross-susceptibility of the Stokes photon induced by the anti-Stokes photon, and $\xi(\delta)$ is the self-susceptibility of the Stokes photon. Considering the Doppler-broadened medium and the all-copropagating scheme, the self-susceptibility and cross-susceptibility are given by
\begin{eqnarray}
\label{eq:self_chi}
	\frac{k_s L}{4} \xi(\delta) \!\!\! &=& \!\!\! \frac{\alpha_s\Gamma_3}{2} \int_{-\infty}^{\infty} d\omega_D  
			\frac{e^{-\omega_D^2/\Gamma_D^2}}{\sqrt{\pi}\Gamma_D}
		\\	&\times&				
		\frac{\delta+i\gamma}{\Omega_c^2-4(\delta+i\gamma)(\delta+\omega_D+i\Gamma_3/2)},
		\nonumber				
		\\
\label{eq:cross_chi} 
	\frac{\sqrt{k_{as} k_s}L}{2} \chi(\delta) \!\!\! &=& \!\!\!
		\frac{\sqrt{\alpha_{as}\alpha_s}\sqrt{\Gamma_3\Gamma_4}}{4}  \int_{-\infty}^{\infty} d\omega_D  
			\frac{e^{-\omega_D^2/\Gamma_D^2}}{\sqrt{\pi}\Gamma_D}
		\\	&\times&				
		\frac{\Omega_p}{\Delta_p-\omega_D + i\Gamma_4/2} 
		\nonumber \\	&\times&	
		\frac{\Omega_c}{\Omega_c^2-4(\delta+i\gamma)(\delta+\omega_D+i\Gamma_3/2)},~~ 
		\nonumber				
\end{eqnarray}
where $\omega_D$ is the Doppler shift, $\Gamma_D$ is the Doppler width, $\alpha_s = n \sigma_s L$ ($n$ is the atomic density and $\sigma_s$ is the resonant absorption cross section of the Stokes transition) is the optical depth of the entire atoms interacting with the Stokes photon on resonance, $\alpha_{as}$ means the similar optical depth of the anti-Stokes transition, $\Omega_p$ and $\Omega_c$ are the Rabi frequencies of the pump and coupling fields, $\Gamma_3$ and $\Gamma_4$ are the spontaneous decay rates of the excited states (i.e., $|3\rangle$ and $|4\rangle$ in Fig.~\ref{fig:transition_diagram}) in the Stokes and anti-Stokes transitions, respectively, $\Delta_p$ is the detuning of the pump field, and $\gamma$ is the dephasing rate of the ground-state coherence, i.e., the decoherence rate. Since the difference between $\Gamma_3$ and $\Gamma_4$ is merely about 5\% in our case, we neglect the difference and set $\Gamma_3 = \Gamma_4 \equiv \Gamma = 2\pi\times6~{\rm MHz}$ in this work. Due to the temperature of the vapor cell being 57 $^{\circ}$C, $\Gamma_D =$ 54$\Gamma$ in the calculation~\cite{OurSR2018}.

In the biphoton's spectral profile shown in Eq.~(\ref{eq:biphoton_frequency}), we first derive the analytical expression corresponding to the EIT effect, i.e.,
\begin{eqnarray}
	T_{\rm EIT} \!\!&=&\!\! 
		\left| \exp\!\! \left[ i\frac{k_s L}{4} \xi(\delta) \right] \right|^2
		\nonumber \\
	\!\!&=&\!\!
		\exp\!\! \left[ -\alpha_s' \!\! \int_{-\infty}^{\infty} \!\! d\omega_D  
		\:e^{-\omega_D^2/\Gamma_D^2} 
		\frac{\beta}{(\omega_D-\omega_0)^2+\beta^2} \right],~~
\label{eq:T_exact}
\end{eqnarray}
where
\begin{eqnarray}
\label{eq:alpha_s_p}
    \alpha_s'&=&\alpha_s  \frac{\sqrt{\pi}\Gamma}{4\Gamma_D} \\
\label{eq:x_0}
	\omega_0 &=& 
		\frac{\delta \Omega_c^2}{4(\delta^2+\gamma^2)} -\delta, \\
\label{eq:beta}		
	\beta &=& 
		\frac{2 \delta^2 \Gamma+\gamma(2\gamma\Gamma+\Omega_c^2)}
		{4(\delta^2+\gamma^2)} 
		\approx \frac{2 \delta^2 \Gamma+\gamma\Omega_c^2}{4(\delta^2+\gamma^2)}.
\end{eqnarray}
The above approximation is valid under the assumption of $2\gamma\Gamma \ll \Omega_c^2$, which is very reasonable in typical EIT experiments. We name $T_{\rm EIT}$ as the biphoton's EIT spectrum.

In Eq.~(\ref{eq:T_exact}), the integral results in the Voigt function, i.e., the convolution between a Gaussian and a Lorentzian functions. The Voigt function can be expressed as the real part of the Faddeeva function, $w(z)$. Thus,
\begin{equation}
	T_{\rm EIT} = \exp \left\{ - \alpha_s' {\rm Re}[w(z)] \right\},
\end{equation}
where $w(z)\equiv {\rm exp}(-z^2) {\rm erfc}(-i z)$ and $z=(\omega_0+i \beta)/\Gamma_D$. Re[$w(z)$] can be approximated as a Lorentzian function of the two-photon detuning ($\delta$) given by
\begin{equation}
	{\rm Re}[w(z)] \approx R\left(1-\frac{A}{1+4\delta^2/\Gamma_L^2}\right),
\label{eq:F_approximation}
\end{equation}
where 
\begin{eqnarray}
\label{eq:R}
	R &=&  \:\exp\! \left( \frac{1}{4 \Gamma_D^2} \right)
		\:{\rm erfc}\! \left( \frac{1}{2  \Gamma_D} \right), \\
\label{eq:A}
	A &=& 1-\frac{1}{R} \:\exp\! \left( \frac{\Omega_c^4}{16 \gamma^2 \Gamma_D^2} \right)
		\:{\rm erfc}\! \left( \frac{\Omega_c^2}{4 \gamma \Gamma_D} \right), \\
\label{eq:linewidth}		
	\Gamma_L &=& 2\gamma 
		\left( 1 +\frac{\Omega_c^2}{4\gamma \Gamma_D} \right).
\end{eqnarray}
In Eq.~(\ref{eq:F_approximation}), $A/(1+4\delta^2/\Gamma_L^2)$ is the Lorentzian function and $\Gamma_L$ is its full width at half maximum (FWHM). Denote the Lorentzian function as $f(\delta)$ and $1-{\rm Re}[w(z)]/R$ as $g(\delta)$. The percentage difference between them, i.e., $\int (f- g)^2 d\delta / \int (f \cdot g) d\delta$, is always less than 5\% as long as $\gamma >$ 0.01$\Gamma$. When $\Omega_c^2/(4\gamma \Gamma_D) \gg 1$, Eq.~(\ref{eq:F_approximation}) becomes not a good approximation. Using Eq.~(\ref{eq:F_approximation}), we obtain the biphoton's EIT spectrum in the Doppler-broadened medium as the following:
\begin{eqnarray}
\label{eq:T_approximation}
	T_{\rm EIT} &=& \:\exp\! \left[ -\alpha_s' R
		\left( 1-\frac{A}{1+ 4\delta^2/\Gamma_L^2} \right) \right]. 
\end{eqnarray}
The FWHM, peak height, and baseline in the above equation are exactly the same as those in the equation below.
\begin{eqnarray}
\label{eq:T_approximation2}
	T_{\rm EIT} &\approx& e^{-\alpha_s' R} 
		\left[1+ \frac{ e^{\alpha_s' RA}-1}{1+4\delta^2/\Gamma_{\rm EIT}^2} \right],\\
\label{eq:Gamma_EIT}
	\Gamma_{\rm EIT} &=& 
		\Gamma_L \sqrt{\frac{ \alpha_s' R A }{ \ln[(1+e^{\alpha_s' R A})/2] } -1}.
\end{eqnarray}
The percentage difference between the right-hand sides of Eqs.~(\ref{eq:T_approximation}) and (\ref{eq:T_approximation2}) is always less than 5\%, and a smaller value of $\alpha_s' R A$ makes the difference less. According to Eq.~(\ref{eq:T_approximation2}), the biphoton's EIT spectrum is approximated as a Lorentzian peak on top of a baseline \cite{OurSR2018}. 

\FigTwo

We compare the biphoton's EIT spectrum calculated from the numerical integral of Eq.~(\ref{eq:T_exact}) with that calculated from the analytical formula of Eq.~(\ref{eq:T_approximation}). Figure~\ref{fig:eit_simulation}(a) shows the comparisons at different values of the coupling Rabi frequency, $\Omega_c$. As $\Omega_c$ becomes larger, i.e., a larger value of $\Omega_c^2/(4\gamma\Gamma_D)$, the deviation between numerical and analytical spectra becomes larger. Figure~\ref{fig:eit_simulation}(b) shows the percentage difference between the numerical and analytical values of the FWHM of biphoton's EIT spectrum. One can observe the percentage difference increases with the value of $\Omega_c^2/(4 \gamma \Gamma_D)$.

We next derive the four-wave mixing (FWM) effect in the biphoton spectral profile shown in Eq.~(\ref{eq:biphoton_frequency}), named the biphoton's FWM spectrum thereafter. The spectrum of FWM transmission is the square of absolute value of Eq.~(\ref{eq:cross_chi}),
\begin{eqnarray}
\label{eq:FWM_part}
	T_{\rm FWM}= \left| \frac{\sqrt{k_{as} k_s}L}{2} \chi(\delta) \right|^2. 
\end{eqnarray} 
At a large value of the pump detuning $\Delta_p \gg \Gamma$, which is the typical experimental condition in the biphoton generation, the term $\Omega_p/(\Delta_p-\omega_D + i\Gamma/2 )$ in Eq.~(\ref{eq:cross_chi}) is treated as a constant of $\Omega_p/\Delta_p$. The integration can be approximated as the following complex Lorentz function:
\begin{eqnarray}
\label{eq:FWM_part_th}
	T_{\rm FWM} &=& 
		 \left| \frac{\Gamma\sqrt{\alpha_{as}\alpha_s}}{4}
		\frac{\Omega_p}{\Delta_p} \right. \left. \frac{B}{1-i (2 \delta/\Gamma_{\rm FWM})} \right|^2, \\ 
\label{eq:Gamma_FWM}
	\Gamma_{\rm FWM} &=& 2\gamma 
		\left( 1 +\frac{\Omega_c^2}{4\gamma \Gamma_D} \right) 
		= \Gamma_L, \\
\label{eq:B}
	B &=& \frac{\sqrt{\pi}\Omega_c}{4 \gamma \Gamma_D}
		\:{\rm exp}\! \left( \frac{\Omega_c^4}{16 \gamma^2 \Gamma_D^2} \right) 
		\:{\rm erfc}\! \left( \frac{\Omega_c^2}{4 \gamma \Gamma_D} \right).	
\end{eqnarray} 
The above approximation is valid under the condition of $\Gamma_{\rm FWM} \gg \Omega_c^2/(5.4 \Gamma_D)$, where $\Omega_c^2/(5.4 \Gamma_D)$, obtained numerically, is the separation between the two transmission peaks in the biphoton's FWM spectrum. Thus, the spectral profile of cross-susceptibility is a Lorentzian function with a FWHM of $\Gamma_{\rm FWM}$, which is equal to $\Gamma_L$.

\FigThree

We compare the biphoton's EIT spectrum calculated from the numerical integral of Eq.~(\ref{eq:FWM_part}) with that calculated from the analytical formula of Eq.~(\ref{eq:FWM_part_th}). Figure~\ref{fig:fwm_simulation}(a) shows the numerical and analytical spectra of FWM transmission at the coupling Rabi frequencies $\Omega_c$ of 2.7$\Gamma$, 3.8$\Gamma$ and 5.4$\Gamma$, respectively. The deviation between the numerical and analytical spectra is more prominent as $\Omega_c$ increases, because the separation between the two transmission peaks in the FWM spectrum gets larger due to a larger value of $\Omega_c$. Figure~\ref{fig:fwm_simulation}(b) shows the percentage difference between the numerical and analytical values of the FWHM of biphoton's FWM spectrum. One can observe the difference increases with the value of $\Omega_c^2/(4 \gamma \Gamma_D)$.

\FigFour

The biphoton's overall spectrum shown by Eq.~(\ref{eq:biphoton_frequency}) is approximately equal to the product of the biphoton's EIT spectrum, $T_{\rm EIT}$, and the biphoton's FWM spectrum. $T_{\rm FWM}$. Since $|{\rm sinc} \! \left[ k_s L \xi(\delta) /4 \right]|$ is slowly-varying within the biphoton's spectral width,  Eq.~(\ref{eq:biphoton_frequency}) without ${\rm sinc} \! \left[ k_s L \xi(\delta) /4 \right]$ changes the result little. As shown in Eq.~(\ref{eq:T_approximation2}), the analytical expression of biphoton's EIT spectrum is proportional to 
\begin{equation}
\label{eq:eit_bi}
	1+ \frac{\sigma-1}{1+ 4\delta^2/\Gamma_{\rm EIT}^2}, 
\end{equation}
where $\sigma = e^{\alpha'_s R A}$ is a constant independent of $\delta$. As shown in Eq.~(\ref{eq:FWM_part_th}), the analytical expression of the biphoton's FWM spectrum is proportional to
\begin{equation}
\label{eq:fmw_bi}
	\frac{1}{1+4\delta^2/(\eta \Gamma_{\rm EIT})^2},
\end{equation}
where $\eta^{-1} = \sqrt{\alpha'_s R A / \ln[(1+e^{\alpha'_s R A})/2] -1}$ based on Eqs.~(\ref{eq:Gamma_EIT}) and (\ref{eq:Gamma_FWM}), which is independent of $\delta$. As the value of $\alpha'_s R A$ is not large, we have $\sigma \approx \eta^2$, and the product of the biphoton's EIT and FWM spectral profiles gives
\begin{eqnarray}
	&& \left( 1+ \frac{\sigma-1}{1+4\delta^2/\Gamma_{\rm EIT}^2} \right) \times
		\frac{1}{1+4\delta^2/(\eta \Gamma_{\rm EIT})^2} \nonumber \\
	&& ~\approx
		\frac{\eta^2}{1+4\delta^2/\Gamma_{\rm EIT}^2}. 
\end{eqnarray}
As the value of $\alpha'_s R A$ is large, $\eta$ and $\sigma$ become large. A large $\eta$ makes the biphoton's FWM linewidth far greater than the biphoton's EIT linewidth, indicating that the product of the biphoton's EIT and FWM spectral profiles is dominated by the EIT profile. We obtain
\begin{eqnarray}
	&& \left( 1+ \frac{\sigma-1}{1+4\delta^2/\Gamma_{\rm EIT}^2} \right) \times
		\frac{1}{1+4\delta^2/(\eta \Gamma_{\rm EIT})^2} \nonumber \\
	&& ~\approx
		1+ \frac{\sigma-1}{1+4\delta^2/\Gamma_{\rm EIT}^2}
		\approx \frac{\sigma}{1+4\delta^2/\Gamma_{\rm EIT}^2}. 
\end{eqnarray}
The second approximation in the above is due to $\sigma \gg 1$. Thus, the linewidth, $\Gamma_{\rm BI}$, of biphoton's overall spectrum is close to that, $\Gamma_{\rm EIT}$, of the biphoton's EIT spectrum.

In Fig.~\ref{fig:three_linewidth_comparison}, we compare the ratio of $\Gamma_{\rm EIT}$ to $\Gamma_{\rm BI}$ and that of $\Gamma_{\rm FWM}$ to $\Gamma_{\rm BI}$, where $\Gamma_{\rm EIT}$, $\Gamma_{\rm BI}$, and $\Gamma_{\rm FWM}$ are numerically calculated from Eqs.~(\ref{eq:biphoton_frequency}), (\ref{eq:T_exact}), and (\ref{eq:FWM_part}), respectively. The ratios are plotted against $\Omega_c^2$ at various values of the OD ($\alpha_s$) and decoherence rate ($\gamma$).
In all the cases, it is apparent that the linewidth of the biphoton's overall spectrum is nearly the same that of the biphoton's EIT spectrum, i.e.,
\begin{equation}
\label{eq:linewidth_bi}
	\Gamma_{\rm BI} \approx \Gamma_{\rm EIT}.
\end{equation}
The outcome of the comparison between the numerical results shown in Fig.~\ref{fig:three_linewidth_comparison} is consistent with the expectation of the discussion utilizing the analytical expressions in Eqs.~(\ref{eq:eit_bi}) and (\ref{eq:fmw_bi}).

The numerical and analytical results of the biphoton's overall spectrum are compared in Fig.~\ref{fig:biphoton_simulation}(a). The solid lines are numerically calculated from the square of Eq.~(\ref{eq:biphoton_frequency}), and the dashed lines are the results of the analytical formula in Eq.~(\ref{eq:T_approximation2}).
Figure~\ref{fig:biphoton_simulation}(b) shows the percentage difference between the numerical and analytical values of the FWHM of biphoton's overall spectrum. The numerical linewidth is determined by the Lorentzian best fit of the biphoton spectrum calculated from Eq.~(\ref{eq:biphoton_frequency}). The analytical linewidth is given by the formula in Eq.~(\ref{eq:linewidth_bi}) or equivalently Eq.~(\ref{eq:Gamma_EIT}). In Fig.~\ref{fig:biphoton_simulation}(b), the percentage difference increases with the value of $\Omega_c^2/(4 \gamma \Gamma_D)$.

\FigFive

We have derived the analytical formula of the biphoton's self-susceptibility [$(k_s L/4) \xi(\delta)$] or EIT spectrum as shown by Eq.~(\ref{eq:T_approximation2}), and that of the biphoton's cross-susceptibility  [$(\sqrt{k_{as} k_s}L/2) \chi(\delta)$] or FWM spectrum as shown by Eq.~(\ref{eq:FWM_part_th}). In addition, ${\rm sinc} \! \left[ k_s L \xi(\delta) /4 \right]$ in Eq.~(\ref{eq:biphoton_frequency}) plays little role in the spectral profile. Therefore, the analytical expression of Eq.~(\ref{eq:biphoton_frequency}) is obtained, and the time-correlation function or the wave packet of biphotons in Eq.~(\ref{eq:biphoton}) becomes
\begin{eqnarray}
\label{eq:g2_approx}
	G^{(2)}(\tau) &=&  
		C e^{-\Gamma_{\rm BI} t}, \\
\label{eq:C_value}
	C & \propto & (\alpha_{as}\alpha_s \Gamma^2) \frac{\Omega_p^2}{\Delta_p^2} B^2 e^{-\alpha_s'R(1-A)}.
\end{eqnarray} 
The biphoton's wave packet is an exponential-decay function with the decay time constant of $1/\Gamma_{\rm BI}$. As long as the value of $\Omega_c^2/(4\gamma \Gamma_D)$ is not far greater than 1, Eq.~(\ref{eq:g2_approx}) is a reasonable approximation.
\section{Experimental Setup}

We experimentally studied the temporal width and spectral linewidth of SFWM biphotons generated from a hot vapor of $^{87}$Rb atoms. The SFWM transition scheme is shown in Fig.~\ref{fig:transition_diagram}, and the actual energy levels in the experiment are specified in the caption. The time correlation function (i.e., the wave packet) of biphotons and the corresponding EIT spectrum were measured as functions of the coupling power or the square of the coupling Rabi frequency $\Omega_c^2$. In the measurement of the EIT spectrum, a weak probe laser field was employed in additional to the coupling field, and its Rabi frequency is far less than $\Omega_c$ to satisfy the perturbation limit and to affect the ground-state population distribution very little. The two laser fields form the $\Lambda$-type transition scheme with zero one-photon detuning, and their transitions are depicted in Fig.~1(a). Both laser fields have the wavelengths of about 795 nm. The probe laser was injection-locked by the coupling laser with an offset frequency, which was provided by an fiber-based electro-optic modulator (EOM) with an input RF frequency of around 6835 MHz. The injection-lock scheme ensures the two-photon detuning, $\delta$, between the probe and coupling fields to be stable and free of the laser frequency fluctuations. We swept the RF frequency of the EOM, i.e., the probe frequency which is equal to the two-photon detuning, and measured the probe transmission in the spectroscopic measurement.

The coupling and input probe fields completely overlapped and propagated through the same direction in the measurement of the EIT spectrum, minimizing the decoherence rate induced by the Doppler effect. The two fields had the $s$ and $p$ polarizations, and the $e^{-2}$ full widths of 1.5 and 1.4 mm. The output probe beam was collected by a polarization-maintained fiber (PMF), and its power after the PMF was detected by a Thorlabs PDA36A photo detector. The PMF's collection efficiency for the probe field is 70\%. We produced the coupling (or pump) field from an external-cavity diode laser of Toptica DL DLC pro 795 (or 780). A homemade 795 nm bare-diode laser under the injection lock generated the probe field. The coupling frequency was stabilized by the saturated absorption spectroscopy, and the pump frequency was stabilized by a wave meter.

In the measurement of the biphoton wave packet or two-photon correlation function as a function of the delay time between the photon pair, we employed only the coupling and pump fields. The coupling field is the same as that used in the measurements of the EIT spectrum. The pump field had the $p$ polarization. We set the pump detuning to $2$ GHz. The $e^{-2}$ full width of the pump field is 1.2 mm. Pairs of anti-Stokes (or signal) and Stokes (or probe) photons were generated by the spontaneous four-wave mixing (SFWM) process. The coupling and pump laser fields and the signal and Stokes probe photons all propagated in the same direction. This all-copropagation scheme can satisfy the phase match condition. The biphotons or the pairs of anti-Stokes and Stokes photons were collected by two PMFs, and their counts after the PMFs detected by two single-photon counting modules (SPCMs). To block the leakages of the strong coupling and pump fields to the SPCMs, we utilized polarization filters and etalon filters which totally provide the extinction ratios of about 130 dB. 

We applied a hyperfine optical pumping (HOP) field to optically pump the population out of the state of $|5S_{1/2}, F=1\rangle$ during the measurements of the biphoton wave packet and the EIT spectrum. The HOP field has a hollow-shaped beam profile, while the interaction region of the SFWM process and the EIT spectrum is in its hollow region. We set the power of the HOP field to 9 mW and stabilized its frequency to 80 MHz below the transition frequency of $|5S_{1/2},F=1\rangle$ $\rightarrow$ $|5P_{3/2},F=2\rangle$. The HOP field increased decoherence rate a little in the experimental system. Other details of the experimental setup and measurement method of the biphoton wave packet can be found in our previous works \cite{Ourhotbiphoton1, Ourhotbiphoton2}.

\section{Results and Discussion} \label{sec:results}

We systematically measured the EIT spectra and the biphoton wave packets. The temperature of the vapor cell, which consists of nearly only $^{87}$Rb atoms, was maintained at 57 $^{\circ}$C throughout the experiment. At this temperature, $\Gamma_D =$ 54$\Gamma$. We utilized the absorption spectrum of a weak probe field in the presence of the coupling and HOP fields to determine the value of $\alpha_s$ (OD). In the measurement of the absorption spectrum, the probe frequency was swept across the entire Doppler-broadened spectral lines of $|5S_{1/2},F=2\rangle$ $\rightarrow$ $|5P_{1/2},F=1\rangle$ and $|5S_{1/2},F=2\rangle$ $\rightarrow$ $|5P_{1/2},F=2\rangle$. We fitted the absorption spectrum and determined $\alpha_s =$ 360$\pm$10.

In the measurement of biphoton wave packets, we used five different coupling powers of 0.5, 1, 2, 4, and 8 mW. The coupling Rabi frequency was determined by the EIT spectrum, and the determination method will be explained in the next paragraph. The pump power or Rabi frequency was fixed to 0.5 mW or approximately 2.0$\Gamma$, which is estimated according to the peak intensity of the pump beam. In the measurement of EIT spectra, we set the probe power to 0.5 $\mu$W. The probe field was weak enough that it can be treated as the perturbation. The coupling power in each EIT spectrum was the same as that in the corresponding biphoton wave packet.  

\FigSix

Figures~\ref{fig:experiment_data}(a)-(e) show the EIT spectra at various coupling powers. The blue lines represent the experimental data. The red lines are the theoretical predictions calculated from  with Eq.~(\ref{eq:T_exact}). In the theoretical calculation, we set the OD ($\alpha_s$) to 360 as determined by the absorption spectrum, and kept the ratio between the five coupling Rabi frequencies ($\Omega_c$) as the square root of the ratio between the five coupling powers, i.e., the Rabi frequencies correspond to $\Omega_{c0}$, $\sqrt{2}\Omega_{c0}$, $2\Omega_{c0}$,  $2\sqrt{2}\Omega_{c0}$, and  $4\Omega_{c0}$. We tuned a single value of $\Omega_{c0}$ for all EIT spectra, and varied the value of the decoherence rate ($\gamma$) for each EIT spectrum to get a good match between the data and the predictions.  The value of $\Omega_{c0}$ used in the calculation was 1.75$\Gamma$, while the Rabi frequency estimated from the peak intensity of the 0.5-mW coupling power was 2.05$\Gamma$. When we boosted the coupling power by 16 folds, the value of $\gamma$ increased by a factor of 4.3. The change of the biphoton's temporal width was mainly influenced by the coupling power. The consistency between the experimental data and the theoretical predictions is satisfactory.

The green dashed lines in Figs.~\ref{fig:experiment_data}(a)-(e) are the best fits of Lorentzian functions, since the EIT spectral profile is approximately a Lorentzian function as shown by Eq.~(\ref{eq:T_approximation2}). The measured EIT spectral profile of a Doppler-broadened medium is very close to a Lorentzian function. Please note that the biphoton's EIT spectrum is given by $|\exp[i(k_s L/4)\xi]|^2$ as shown in Eq.~(\ref{eq:T_exact}), and the EIT spectrum with an input probe field is given by $|\exp[i(k_s L/2) \xi]|^2$. The difference of a factor of 2 between the two expressions is as a matter of fact that the probe photons resulted from the SFWM process are generated everywhere in a medium, and they propagate through a half of the medium length on average. To apply Eq.~(\ref{eq:T_approximation}) or (\ref{eq:T_approximation2}) to the EIT spectrum with an input probe field, $\alpha_s'$ in the equation needs to be redefined as $\alpha_s \sqrt{\pi}\Gamma/(2\Gamma_D)$, which differs from Eq.~(\ref{eq:alpha_s_p}) by a factor of 2.

\FigSeven

The blue lines in Figs.~\ref{fig:experiment_data}(f)-(j) are the experimental data of biphoton wave packets taken at the same coupling powers as those in Figs.~\ref{fig:experiment_data}(a)-(e), respectively. The red lines are the theoretical predictions calculated from Eq.~(\ref{eq:biphoton}). We set the calculation parameters of $\Omega_c$, $\alpha_s$, and $\gamma$ to the values determined by the absorption and EIT spectra. A Lorentzian function with a linewidth of 35 MHz, corresponding to the spectral profile of the etalon filter used in the experiment, was added to Eq.~(\ref{eq:biphoton_frequency}) in the calculation. Since the biphoton wave packets have the linewidths between 270 kHz and 2.2 MHz, the etalon filter's linewidth affects the calculation results little. The green dashed lines are the best fits of exponential-decay functions, i.e., $y(t) = y_0+A \times{\rm exp} [-(t-t0)/\tau]$, since the biphoton wave packet behaves approximately like an exponential-decay function as shown by Eq.~(\ref{eq:g2_approx}). We first determined the value of $y_0$ from the baseline of data, then fixed the value of $t_0$, and finally fitted the values of $A$ and $\tau$. The theoretical predictions and the exponential-decay best fits are in good agreement with the experimental data. In Fig.~\ref{fig:experiment_data}, we have demonstrated that the EIT spectrum measured with classical light can be a good indicator of the biphoton wave packet generated by the SFWM process.

To verify whether the temporal width of the measured biphoton wave packet is consistent with the analytical formula of $1/\Gamma_{\rm BI}$, we plot the biphoton's temporal width as a function of the coupling power in Fig.~\ref{fig:eit_biphoton_linewidth}. The red circles are the biphoton's temporal widths obtained from the $e^{-1}$ time constants of the best fits as the examples shown in Figs.~\ref{fig:experiment_data}(f)-(j). Each data point is the average of repeated measurements. As for the analytical formulas of $1/\Gamma_{\rm BI}$, the blue squares are the results calculated from Eq.~(\ref{eq:Gamma_EIT}) or (\ref{eq:linewidth_bi}). The calculation parameters of $\Omega_c$, $\alpha_s$, and $\gamma$ are the values determined in Figs.~\ref{fig:experiment_data}(a)-(e). Since the values of $\Omega_c^2/ (4\gamma\Gamma_D)$ of the five coupling powers ranged between 0.71 and 2.6, Eq.~(\ref{eq:g2_approx}) as well as the formula of $1/\Gamma_{\rm BI}$ is a good approximation. As expected, the agreement between the data of the measured temporal width and the predictions of the analytical formula is satisfactory.

\section{Conclusion}

We have systematically studied the temporal profile of biphotons generated from a hot atomic vapor via the SFWM. Taking into account all velocity groups in a Doppler-broadened medium, we were able to derive the analytical formula of the biphoton wave packet or the two-photon correlation function. The derived formula shows that the biphoton's spectral profile is a Lorentzian function with the linewidth mainly determined by the EIT linewidth, $\Gamma_{\rm EIT}$, where the formula of $\Gamma_{\rm EIT}$ is given by Eq.~(\ref{eq:Gamma_EIT}). Thus, the biphoton wave packet is close to an exponential-decay function with the $e^{-1}$ decay time constant of $1/\Gamma_{\rm EIT}$. 

Since the coupling power can significantly affect the EIT effect and the biphoton's temporal width, we measured the EIT spectra with classical light and the biphoton wave packets generated from the SFWM process at various coupling powers from 0.5 to 8 mW. The measured EIT spectrum exhibited a Lorentzian profile, which is consistent with the prediction of the analytical formula shown in Eq.~(\ref{eq:T_approximation2}). Under the same experimental condition as the EIT spectrum, the measured biphoton wave packet behaved mainly like an exponential-decay function, in agreement with the analytical formula shown in Eq.~(\ref{eq:g2_approx}). We utilized the experimental parameters determined by the EIT spectra to calculate the theoretical predictions of the biphoton wave packets. The agreement between the theoretical predictions and experimental data is satisfactory. Furthermore, we compared the biphoton temporal width obtained from the measurement with that calculated from the analytical formula given by   Eq.~(\ref{eq:Gamma_EIT}) or (\ref{eq:linewidth_bi}). The comparison verifies the analytical formula is a reasonable approximation as long as the value of $\Omega_c^2/(4\gamma \Gamma_D)$ is not far greater than 1.

In conclusion, the EIT spectrum measured with classical light is a good indicator for the biphoton wave packet generated by the SFWM process. The analytical formulas derived in this work provide a simple way to quantitatively understand the temporal and spectral profiles of the biphotons .

\section*{Acknowledgments}
This work was supported by Grant Nos.~109-2639-M-007-001-ASP and 110-2639-M-007-001-ASP of the Ministry of Science and Technology, Taiwan.


\end{document}